\documentstyle[12pt]{article}
\newlength{\extraspace}
\newlength{\extraspaces}
\newcommand{\be}{\begin{equation}
\addtolength{\abovedisplayskip}{\extraspaces}
\addtolength{\belowdisplayskip}{\extraspaces}
\addtolength{\abovedisplayshortskip}{\extraspace}
\addtolength{\belowdisplayshortskip}{\extraspace}}
\newcommand{\ee}{\end{equation}}

\newcommand{\ba}{\begin{eqnarray}
\addtolength{\abovedisplayskip}{\extraspaces}
\addtolength{\belowdisplayskip}{\extraspaces}
\addtolength{\abovedisplayshortskip}{\extraspace}
\addtolength{\belowdisplayshortskip}{\extraspace}}
\newcommand{\ea}{\end{eqnarray}}

\newcommand{\nonu}{\nonumber \\[.5mm]}
\newcommand{\A}{&\!\!\!}

\newcommand{\newsection}[1]{
\vspace{7mm} \pagebreak[3] \addtocounter{section}{1}
\setcounter{subsection}{0} \setcounter{footnote}{0}
\begin{center}
{\large {\bf \thesection. #1}}
\end{center}
\nopagebreak
\medskip
\nopagebreak \hspace{2mm}}

\begin{document}

\begin{center}
{{\bf Energy distribution of Kerr spacetime using M\o ller energy
momentum complex }}
\end{center}
\centerline{ Gamal G.L. Nashed}

\bigskip

\centerline{{\it Mathematics Department, Faculty of Science, Ain
Shams University, Cairo, Egypt }}

\bigskip
 \centerline{ e-mail:nashed@asunet.shams.edu.eg}
\hspace{2cm}
\\
\\
\\
\\
\\
\\
\\
\\
\\
\\

 Using the energy momentum complex given by M\o ller in 1978 based on the
 absolute parallelism, the energy distribution in Kerr spacetime is evaluated.
 The energy with this spacetime is found to be the same as it was earlier evaluated
  using different definitions mainly based on the metric tensor.

\newpage
\begin{center}
\newsection{\bf Introduction}
\end{center}

After Einstein's original pseudotensor, a large number of
expressions for the energy distribution in a general relativistic
system have been proposed by many authors \cite{Tr}$\sim$
\cite{KO}. To get a meaningful results for the energy in the
prescription of Einstein, Tolman or Landau and Lifshitz one is
compelled to use a quasi-Galilean\footnote{By a quasi-Galilean
coordinate we mean a coordinate in which the metric tensor $g_{\mu
\nu}$ approaches the Minkowski spacetime metric $\eta_{\mu
\nu}=diag(1,-1,-1,-1)$ at spatial infinity.}. M\o ller revived
\cite{Mc} the issue of energy and momentum in general
 relativity, and required \cite{Mc1} that any energy-momentum complex
 ${\tau_{\mu}}^{\nu}$ must satisfy the following properties:
(A) It must be an affine tensor density which satisfies
conservation law, (B) for an isolated  system the quantities
 $P_{\mu}$ are constant in time and transform as the covariant components
 of a 4-vector under linear coordinate transformations,
  and  (C) the superpotential  ${{\cal U_{\mu}}^{\nu \lambda}}$ $=-
  {{\cal U_{\mu}}^{\lambda \nu}}$ transforms as a tensor density of rank 3
  under the group of spacetime transformations.

It is not possible to satisfy all the above requirements if the
gravitational field is described by the metric tensor alone
\cite{Mc2}. In a series of papers \cite{Mc2,Mc3,Mc4} therefore,
M\o ller was led to the tetrad description of gravitation, and
constructed a formal form of energy-momentum complex that
satisfies all the requirements. The metric tensor is uniquely
fixed by the tetrad field, but the reverse is not true, since the
tetrad has six extra degrees of freedom. In the tetrad formulation
of general relativity, the tetrad field is allowed to undergo
local Lorentz transformations with six arbitrary functions. The
energy-momentum complex is not a tensor and changes
 its form under such  transformations. Therefore, unless one can find a
 good physical argument for fixing the tetrad throughout the system, one
 cannot speak about the energy distribution inside the system. The total
 energy-momentum obtained by the complex, however, is invariant under local
 Lorentz transformations with appropriate boundary conditions
 \cite{Mc4,Mo1}.

Dymnikova \cite{Di} derived a static spherically symmetric
nonsingular black hole solution in orthodox general relativity
assuming a specific form of the stress-energy momentum tensor.
This solution practically coincides with the Schwarzschild
solution for large $r$,  for small $r$ it behaves like the de
Sitter solution and describes a spherically symmetric black hole
singularity free everywhere \cite{Di}. The calculations of the
 energy associated with this metric was done by  Radinschi
 \cite{Ra}.

 Nashed \cite{Ng} has obtained two general spherically symmetric non
 singular black hole solutions in M\o ller's tetrad theory of
 gravitation. One of those solutions is characterized by an
 arbitrary function while the other by a constant. The associated
 metric of these solutions are the same and gave the metric
 obtained before by Dymnikova \cite{Di}. The energy content of
 those two solutions are calculated using the energy momentum
 complex given by M\o ller. It is shown that the energy of the two
 solutions depends on the arbitrary function and on the constant
 respectively and they are different from each other and also different from
 the energy given before  by Radinschi \cite{Ra}. Nashed \cite{Ng1}
  also show that the calculations of energy using the prescription
  of M\o ller in the framework of absolute parallelism spacetime are more
  accurate than that given in the framework of the Riemannian
  spacetime.Toma \cite{Tn} gives an exact solution to the vacuum field equation of
the new general relativity.

  It is the aim of the present work to calculate the energy
  distribution  of the Kerr spacetime given by Toma \cite{Tn}
   using the energy momentum complex
  given by M\o ller. In section 2 we briefly review the new
  general relativity theory of gravitation. The Kerr solution obtained by
Toma \cite{Tn} and its energy distribution  are given in section
3. Section 4 is devoted to discussion.

\newsection{The new general relativity theory of gravitation}

The fundamental fields of gravitation are the parallel vector
fields ${b_k}^\mu$. The component of the metric tensor $g_{\mu
\nu}$ are related to the dual components ${b^k}_\mu$ of the
parallel vector fields by the relation \be g_{\mu \nu}=
{b^i}_\mu{b^i}_\nu. \ee
  The nonsymmetric connection
 ${\Gamma^\lambda}_{\mu \nu}$\footnote{Latin indices $(i,j,k,\cdots)$ designate the vector
number, which runs from $(0)$ to
 $(3)$, while Greek indices $(\mu,\nu,\rho, \cdots)$ designate the world-vector components
running from 0 to 3. The spatial part of Latin indices is denoted
by $(a,b,c,\cdots)$, while that of Greek indices by $(\alpha,
\beta,\gamma,\cdots)$.} are defined by
 \be
{\Gamma^\lambda}_{\mu \nu} ={b_k}^\lambda {b^k}_{\mu,\nu}, \ee as
a result of the absolute parallelism \cite{HS}.

The gravitational Lagrangian ${\it L}$ of this theory is an
invariant constructed from the quadratic terms of the torsion
tensor
 \be
 {T^\lambda}_{\mu \nu} \stackrel{\rm def.}{=}{\Gamma^\lambda}_{\mu
 \nu}-{\Gamma^\lambda}_{\nu \mu}. \ee
  The following Lagrangian
\be {\cal L} \stackrel{\rm def.}{=} -{1 \over 3\kappa} \left(
t^{\mu \nu \lambda} t_{\mu \nu \lambda}- v^\mu v_\mu \right)+\zeta
a^\mu a_\mu, \ee is quite favorable experimentally \cite{HS}. Here
$\zeta$ is a constant parameter,  $\kappa$ is the Einstein
gravitational constant and $t_{\mu \nu \lambda}, v_\mu$ and
$a_\mu$ are the irreducible components of the torsion tensor:

\ba t_{\lambda \mu \nu}\A=\A {1 \over 2} \left(T_{\lambda \mu
\nu}+T_{\mu \lambda \nu} \right) +{1 \over 6} \left( g_{\nu
\lambda}V_\mu+g_{\mu \nu}V_\lambda \right) -{1 \over 3} g_{\lambda
\mu}V_\nu,\nonu
V_\mu \A=\A {T^\lambda}_{\lambda \nu}, \nonu
a_\mu \A=\A {1 \over 6} \epsilon_{\mu \nu \rho \sigma}T^{\nu \rho
\sigma},
 \ea
where $\epsilon_{\mu \nu \rho \sigma}$ is defined by \be
\epsilon_{\mu \nu \rho \sigma} \stackrel{\rm def.}{=} \sqrt{-g}
 \delta_{\mu \nu \rho \sigma} \ee with $\delta_{\mu \nu \rho
\sigma}$ being completely antisymmetric and normalized as
$\delta_{0123}=-1$.

By applying the variational principle to the Lagrangian (4), the
gravitational field equation are given by \cite{HS}\footnote{We
will denote the symmetric part by ( \ ), for example, $A_{(\mu
\nu)}=(1/2)( A_{\mu \nu}+A_{\nu \mu})$ and the  antisymmetric part
by the square bracket [\ ], $A_{[\mu \nu]}=(1/2)( A_{\mu
\nu}-A_{\nu \mu})$ .}: \be G_{\mu \nu} +K_{\mu \nu} = -{\kappa}
T_{(\mu \nu)}, \ee \be {b^i}_\mu{b^j}_\nu
\partial_\lambda(\sqrt{-g} {J_{i j}}^{\lambda})=\lambda
\sqrt{-g}T_{[\mu \nu]}, \ee where the Einstein tensor $G_{\mu
\nu}$ is defined by \be G_{\mu \nu}=R_{\mu \nu}-{1 \over 2} g_{\mu
\nu} R, \ee \be {R^\rho}_{\sigma \mu \nu}=\partial_\mu \left
\{_{\sigma \nu}^\rho \right\}-\partial_\nu \left \{_{\sigma
\mu}^\rho \right\}+\left \{_{\tau \mu}^\rho \right\} \left
\{_{\sigma \nu}^\tau \right\}-\left \{_{\sigma \mu}^\tau \right\}
\left \{_{\tau \nu}^\rho \right\}, \ee \be R_{\mu
\nu}={R^\rho}_{\mu \rho \nu}, \ee \be R=g^{\mu \nu}R_{\mu \nu},
\ee and $T_{\mu \nu}$ is the energy-momentum tensor of a source
field of the Lagrangian $L_m$ \be T^{\mu \nu} = {1 \over
\sqrt{-g}} {\delta {\cal L}_M \over \delta {b^k}_\nu} b^{k \mu}
\ee with ${L_M}= {{\cal L}_M/\sqrt{-g}}$. The tensors $K_{\mu
\nu}$ and $J_{i j \mu}$ are defined as \be K_{\mu \nu}={\kappa
\over \lambda}\left( {1 \over 2} \left[{\epsilon_\mu}^{\rho \sigma
\lambda}(T_{\nu \rho \sigma}-T_{\rho \sigma
\nu})+{\epsilon_\nu}^{\rho \sigma \lambda}(T_{\mu \rho
\sigma}-T_{\rho \sigma \mu}) \right]a_\lambda-{3 \over 2} a_\mu
a_\nu-{3 \over 4}g_{\mu \nu} a^\lambda a_\lambda \right), \ee \be
J_{i j \mu}=-{3 \over 2} {b_i}^\rho {b_j}^\sigma \epsilon_{\rho
\sigma \mu \nu} a^\nu, \ee respectively. The dimensionless
parameter $\lambda$ is defined by \be {1 \over \lambda}={4 \over
9} \zeta+{1 \over 3 \kappa}.\ee In this paper we are going to
consider the vacuum gravitational field: \be T_{(\mu \nu)}=T_{[\mu
\nu]}=0.\ee

\newsection{Kerr Solution and its energy contents}
Toma \cite{Tn} has obtained a solution in the new general
relativity which gave the Kerr metric. The covariant form of the
parallel vector field of this solution is given in the
Boyer-Lindquist coordinate $(t, \rho,\theta,\phi)$ by \ba b_{0
\scriptstyle{0}} \A= \A (1-\displaystyle{a \rho \over 2 \Sigma}),
\qquad \qquad b_{0 \scriptstyle{1}} =  \displaystyle{a \rho \over
2 \Delta}, \nonu
b_{0 \scriptstyle{2}} \A= \A 0, \qquad \qquad \qquad \qquad b_{0
\scriptstyle{3}} = -\displaystyle{a \rho h \sin^2 \theta \over 2
\Sigma}, \nonu
b_{1 \scriptstyle{0}} \A= \A  i\displaystyle{a \rho \sin \theta
\cos \Phi \over 2 \Sigma}, \qquad \quad b_{1 \scriptstyle{1}} = i
\displaystyle{\rho \sin \theta \over \Delta} ( X-\displaystyle{a
\cos \Phi \over 2}),\nonu
b_{1 \scriptstyle{2}} \A= \A i X \cos \theta, \qquad \qquad b_{1
\scriptstyle{3}} = i(\displaystyle{a \rho
 h \sin^3 \theta \cos \Phi \over 2 \Sigma}-Y \sin \theta),\nonu
b_{2 \scriptstyle{0}} \A= \A  i\displaystyle{a \rho \sin \theta
\sin \Phi \over 2 \Sigma}, \qquad b_{2 \scriptstyle{1}} = i
\displaystyle{\rho \sin \theta \over \Delta} ( Y-\displaystyle{a
\sin \Phi \over 2}),\nonu
b_{2 \scriptstyle{2}} \A= \A i Y \cos \theta, \qquad \qquad b_{2
\scriptstyle{3}} =  i(X \sin \theta +\displaystyle{a \rho
 h \sin^3 \theta \sin \Phi \over 2 \Sigma}),\nonu
b_{3 \scriptstyle{0}} \A= \A  i \displaystyle{a \rho \cos \theta
\over 2 \Sigma}, \qquad \qquad b_{3 \scriptstyle{1}} = i (1+
\displaystyle{a \rho \over 2 \Delta}) \cos \theta,\nonu
b_{3 \scriptstyle{2}} \A= \A -i \rho \sin \theta, \qquad \qquad
b_{3 \scriptstyle{3}} = i \displaystyle{a \rho
 h \sin^2 \theta \cos \theta \over 2 \Sigma},
 \ea
 where
 \be
\Sigma=\rho^2+h^2\cos^2 \theta,   \qquad \qquad
\Delta=\rho^2+h^2-a\rho, \ee  \be X=\rho \cos \Phi+h \sin \Phi,
\qquad \qquad Y=\rho \sin \Phi-h\cos \Phi, \ee \be \Phi=\phi-hB,
\qquad \qquad B=\int^\rho \displaystyle{d \rho \over \Delta}. \ee
The parallel vector field (18)  is axially symmetric, i.e., it is
invariant under the transformation \cite{Tn} \ba \A \A
\bar{\phi}\rightarrow \phi+\delta \phi, \qquad \bar{b}_{0 0}
\rightarrow {b}_{0 0}, \qquad  \bar{b}_{11} \rightarrow {b}_{11}
\cos\delta \phi-{b}_{2 2} \sin\delta \phi,\nonu
\A \A   \bar{b}_{2 2} \rightarrow {b}_{2 2} \cos\delta \phi+{b}_{1
1} \sin \delta \phi, \qquad \bar{b}_{33} \rightarrow {b}_{33}. \ea

 The associated metric of the parallel vector field (18) is given by
 \be ds^2=
({1-\displaystyle{a \rho \over \Sigma}}) dt^2 -{\Sigma \over
\Delta} d\rho^2 -\Sigma d\theta^2-\sin^2\theta
\left\{(\rho^2+h^2)+{a\rho h^2\sin^2\theta \over \Sigma} \right
\}d\phi^2-2{a\rho h \sin^2\theta \over \Sigma} dt d\phi, \ee which
is the Kerr metric written in the Boyer-Lindquist coordinates and
a and h are respectively the mass and rotation parameter. This
spacetime has null hypersurface which is given by \be
\rho_{\pm}=\displaystyle{a \pm \sqrt{a^2-4h^2} \over 2}.\ee

There is a ring curvature singularity $\Sigma=0$ in the Kerr
spacetime. This spacetime has an event horizon at $\rho=\rho_+$,
it describes a black hole if and only if $a^2\geq 4h^2$. The
coordinates are singular at $\rho=\rho_{\pm}$. Therefore t is
replaced by a null coordinate \be
dt=d\upsilon-\displaystyle{\rho^2+h^2 \over \Delta}d\rho, \qquad
\qquad d\phi=d\varphi-\displaystyle{h \over \Delta}d\rho,\ee and
the Kerr metric is expressed in the advanced Eddington-Finkelstein
coordinate $(\upsilon, \rho, \theta,\varphi)$ \cite{Xs}  \ba ds^2
\A=\A (1-\displaystyle{a \rho \over \Sigma^2})
d\upsilon^2-2d\upsilon d\rho+{2 a h \rho \sin^2\theta \over
\Sigma^2}d\upsilon d\varphi-\Sigma^2 d\theta^2+2h\sin^2\theta
d\rho d\varphi\nonu
\A \A - \sin^2\theta\left[(\rho^2+h^2)+\displaystyle{a h^2 \rho
\sin^2\theta \over \Sigma^2} \right] d\varphi^2. \ea Transforming
the above to Kerr-Schild Cartesian coordinate (T,x,y,z) according
to \ba T=\upsilon-\rho, \qquad \qquad x=\sin\theta(\rho \cos
\varphi-h \sin \varphi), \nonu
y=\sin\theta(\rho \sin \varphi+h \cos \varphi), \qquad \qquad
z=\rho \cos \theta,\ea the line-element (26) takes the form \be
ds^2 = dT^2-dx^2-dy^2-dz^2-\displaystyle{a\rho^3 \over
\rho^4+h^2z^2}\left(dT+\displaystyle{1 \over \rho^2+h^2} \left[(x
\rho+y h )dx+(y \rho -x h)dy \right]+\displaystyle{z \over \rho}
dz\right)^2.\ee

The superpotential of M\o ller's theory is given by Mikhail et al.
\cite{MWHL} as \be {{\cal U}_\mu}^{\nu \lambda} ={(-g)^{1/2} \over
2 \kappa} {P_{\chi \rho \sigma}}^{\tau \nu \lambda}
\left[\Phi^\rho g^{\sigma \chi} g_{\mu \tau}
 -\lambda g_{\tau \mu} \gamma^{\chi \rho \sigma}
-(1-2 \lambda) g_{\tau \mu} \gamma^{\sigma \rho \chi}\right], \ee
where ${P_{\chi \rho \sigma}}^{\tau \nu \lambda}$ is \be {P_{\chi
\rho \sigma}}^{\tau \nu \lambda} \stackrel{\rm def.}{=}
{{\delta}_\chi}^\tau {g_{\rho \sigma}}^{\nu \lambda}+
{{\delta}_\rho}^\tau {g_{\sigma \chi}}^{\nu \lambda}-
{{\delta}_\sigma}^\tau {g_{\chi \rho}}^{\nu \lambda} \ee with
${g_{\rho \sigma}}^{\nu \lambda}$ being a tensor defined by \be
{g_{\rho \sigma}}^{\nu \lambda} \stackrel{\rm def.}{=}
{\delta_\rho}^\nu {\delta_\sigma}^\lambda- {\delta_\sigma}^\nu
{\delta_\rho}^\lambda. \ee The energy is expressed by the surface
integral \cite{Mc} \be E=\lim_{r \rightarrow
\infty}\int_{r=constant} {{\cal U}_0}^{0 \alpha} n_\alpha dS, \ee
where $n_\alpha$ is the unit 3-vector normal to the surface
element ${\it dS}$.

Now we are in a position to calculate the energy associated with
 the solution (18) using the superpotential (29).
  Thus substituting from (18) into
 (29) we obtain the following nonvanishing values
 \ba
{{\cal U}_0}^{0 1} \A =\A \displaystyle{a \sin\theta \over 16 \pi
(\left((\rho^2+h^2 \cos^2\theta)^2(a \rho-h^2-\rho^2) \right)}(
h^6\cos^2\theta-\rho^2 h^4-\sigma \rho^3 h^2 a \cos^2
\theta+\rho^3 h^2 a \cos^2 \theta \nonu
\A \A -\rho a h^4 \cos^2 \theta -3\rho^3h^2+\rho a h^4 \cos^4
\theta-\rho^4 h^2 \cos^2 \theta-\rho^2 h^4 \cos^4
\theta+2a\rho^5\nonu
\A \A -2\rho^6+\sigma \rho^3 h^2 a+\rho^3 a h^2-h^6\cos^4 \theta)
 \ea \be
{{\cal U}_0}^{0 2}  = \displaystyle{a \rho h^2 \sin^2\theta \cos
\theta \over 16 \pi (\left((\rho^2+h^2 \cos^2\theta)^2(a
\rho-h^2-\rho^2) \right)}( h^2\cos^2\theta+2h^2- 2a \rho -3\rho^2
) \ee \be {{\cal U}_0}^{0 3}  = \displaystyle{h \sin\theta  \over
16 \pi \left( (\rho^2+h^2 \cos^2\theta)^2(a \rho-h^2-\rho^2)
\right)}( 2\rho h^2\cos^2\theta-a h^2\cos^2\theta- 2a \rho^2
+2\rho^3 ). \ee Keeping terms of order a we get \be{{\cal U}_0}^{0
1}  = \displaystyle{ a \sin\theta  \over 16 \pi \left( \rho^2+h^2
\cos^2\theta \right)^2}(\rho^2 h^2\cos^2\theta+
h^4\cos^4\theta-h^4\cos^2\theta+ 2\rho^4 +\rho^2 h^2), \ee
\be{{\cal U}_0}^{0 2}  = \displaystyle{ - a \rho h^2 \sin^2\theta
\cos\theta \over 16 \pi \left( (\rho^2+h^2 \cos^2\theta)^2
(\rho^2+h^2) \right)}(h^2\cos^2\theta+ 2h^2 +3\rho^2),\ee
\be{{\cal U}_0}^{0 3} =\displaystyle{ - \rho h \sin\theta  \over 8
\pi \left(\rho^2+h^2\right)}+ \displaystyle{ a  h^3 \sin\theta
 \over 16 \pi \left( (\rho^2+h^2 \cos^2\theta) (\rho^2+h^2)^2
\right)}(h^2\cos^2\theta+ \rho^2\cos^2\theta +2\rho^2). \ee
 Substituting (36), (37) and (38) into
(32) we get \be E=\displaystyle {a \over 2}.
 \ee
 This result is very satisfactory since if $a=2m$ then (39)
 becomes
 \be
 E=m,\ee as it should be.
 \newsection{Summary}
 The results of the preceding sections can be summarized as
 follows\vspace{.5cm}\\
 1) The exact form of the solutions (18) is  a solution to the
 field equations (7) and (8)  \cite{Tn} and it is also a solution to the field
 equation of Einstein.\vspace{.5cm}\\

 2) The parallel vector field resulting from the transformation
 (27) is an exact solution to the field equations (7) and (8).
 \vspace{.5cm}\\
 3) The energy  distribution of the solution (18) is calculated using
 the energy momentum complex given by M\o ller \cite{Mo1}. Using
 the superpotential (29) we obtained the necessary components
 (36,37,38) required for the calculations of energy. Substituting
 these values of the superpotential in (32) we get the
 satisfactory results $E=m$ as it should be.

\end{document}